# A Grid-Forming HVDC Series Tapping Converter Using Extended Techniques of Flex-LCC

Qianhao Sun, *Member, IEEE*, Ruofan Li, Jichen Wang, Mingchao Xia, *Senior Member, IEEE*, Qifang Chen, *Member, IEEE*, Meiqi Fan, Gen Li, *Senior Member, IEEE*, and Xuebo Qiao, *Member, IEEE*

***Abstract*— This paper discusses an extension technology for the previously proposed Flexible Line-Commutated Converter (Flex-LCC) [1]. The proposed extension involves modifying the arm-internal-electromotive-force control, redesigning the main-circuit parameters, and integrating a low-power coordination strategy. As a result, the Flex-LCC transforms from a grid-forming (GFM) voltage source converter (VSC) based on series-connected LCC and FBMMC into a novel GFM HVDC series tapping converter, referred to as the Extended Flex-LCC (EFLCC). The EFLCC provides dc characteristics resembling those of current source converters (CSCs) and ac characteristics resembling those of GFM VSCs. This makes it easier to integrate relatively small renewable energy sources (RESs) that operate in islanded or weak-grid-supported conditions with an existing LCC-HVDC. Meanwhile, the EFLCC distinguishes itself by requiring fewer full-controlled switches and less energy storage, resulting in lower losses and costs compared to the FBMMC HVDC series tap solution. In particular, the reduced capacity requirement and the wide allowable range of valve-side ac voltages in the FBMMC part facilitate the matching of current-carrying capacities between full-controlled switches and thyristors. The application scenario, system-level analysis, implementation, converter-level operation, and comparison of the EFLCC are presented in detail in this paper. The theoretical analysis is confirmed by experimental and simulation results.**

***Index Terms*— HVDC tap, grid-forming control, islanded or weak-grid-supported RESs, MMC, LCC, HVDC transmission system**

## I. INTRODUCTION

### *A. Technical Background and Application Scenarios*

IN recent decades, the rapid development of power electronics has established High Voltage Direct Current (HVDC) as an integral part of power systems [2]–[4]. Particularly in long-distance and bulk-power transmission applications, HVDC has become an economically attractive option [1], [3] due to its advantages including reduced circuit lines, lower losses, and the ability to connect asynchronous power systems, among others [5]-[7]. Now, two technologies are widely used in the HVDC market: LCC, based on thyristors, and MMC, which uses full-controlled switches [8]-[9]. Both LCC and MMC technologies have reached a high level of maturity and have been widely applied in engineering. For example, in China, the state-of-the-art LCC-HVDC project has a voltage rating of ± 1100 kV, while the world's largest capacity for a single MMC station features a voltage class of ± 800 kV and a capacity of 5000 MW [10]-[11].

However, the implementation of China's dual-carbon goals has brought about some new challenges in HVDC applications:

(1) The LCC-HVDC corridor is required to transmit a higher proportion of clean energy. On one hand, it is possible for the LCC-HVDC corridor to become operational while some of its supporting power systems are still under construction due to construction planning reasons [12]-[13]. This could result in underutilization of the LCC-HVDC if subsequent policy shifts make planned thermal power plants (whose construction has not yet started) incompatible with new policies, thus preventing their construction. On the other hand, the implementation of carbon reduction has increased the emphasis on integrating renewable energy sources (RESs) into power transmission through LCC-HVDC corridors [14]. As a result, some existing LCC-HVDCs need to be upgraded to enable a higher proportion of clean electricity transmission.

(2) The need for very-long-distance HVDC transmission has become increasingly important. There is an inverse correlation between the geographical distribution of regions in China rich in RESs and areas of high energy consumption, similar to the spatial relationship observed between fossil fuel-rich areas and regions with high energy consumption. However, a distinction is that regions abundant in RESs tend to be located further from high energy consumption areas compared to regions developed for thermal power plants, and they cover a larger geographic area. Although the current ± 800 kV LCC and MMC-HVDCs are mature [15]-[16], transmitting over distances of more than 3000km [17] with ± 800 kV HVDCs is still not feasible due to conductor resistances [18]. Furthermore, increasing the voltage level of the HVDC can extend the transmission distance. For instance, a state-of-the-art project in China utilizes ± 1100 kV over 3300 km [10], [18]-[19]. However, this approach presents challenges such as insulation design, device manufacturing, and the large footprint [20], which significantly increase the costs of ± 1100 kV HVDCs. Additionally, increasing the voltage level also has its limitations. For distances beyond 3500km, ± 1100 kV HVDCs also face challenges related to voltage drop and power loss [18] caused by conductor resistances. Therefore, it is crucial to explore new HVDC networking methods for transmitting power over very long distances.

To address the first challenge, one potential solution is to replan the RES power plants in the sending-end grid of the LCC-HVDC corridor. However, it's crucial to note a significant limitation of LCC technology: it cannot connect to weak grids. Consequently, introducing RES-based power plants into the sending-end grid of the LCC-HVDC may not be feasible, as it will bring a decrease in power grid strength. Another alternative [12] is to retrofit the sending-end LCC station while increaing RES-based power plants. This would allow the sending-end station to adapt to the decrease in ac power grid strength. For example, the LCC valve has been modified in [21]-[24] by incorporating controllable capacitors between the LCC valve and the ac transformer. This modification effectively controls the commutation voltage of the LCC valve. Meanwhile, several hybrid topologies at the LCC station level, utilizing MMC as an auxiliary, has been investigated in [1], [25]-[27]. However, modifying the valves or stations in an already built LCC is time-consuming, costly, and challenging due to constraints related to device placement, and valve hall footprint. Besides, planning a sufficient number of RESs near the sending-end regions of the LCC-HVDC also present challenges, and ensuring the safe and stable operation of these RESs, which are connected to the sending-end ac grid, poses additional difficulties.

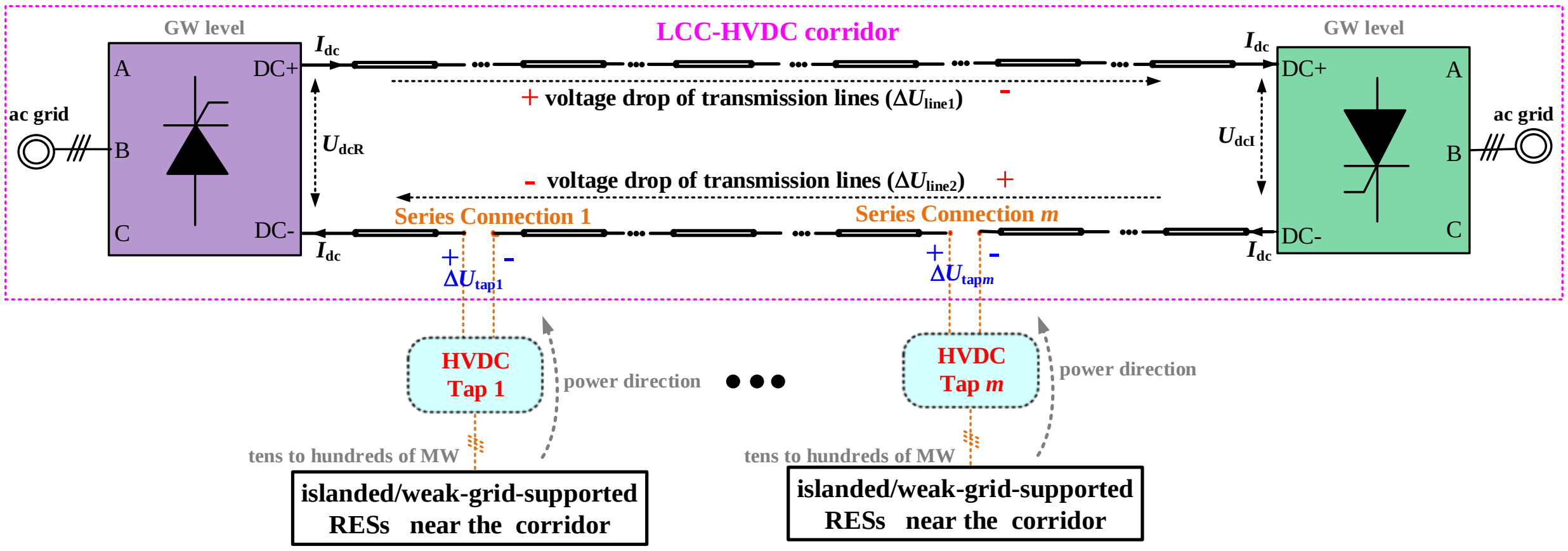

**Fig. 1.** The system-level configuration investigated in this paper.

Comparing the scheme of replanning RESs in the sending-end ac power grid of the LCC-HVDC corridor with the scheme of integrating more RESs adjacent to the LCC-HVDC corridor using HVDC tapping converters [28]-[29], the latter can be a more techno-economic solution. This is due to the LCC-HVDC corridor often spans multiple regions abundant in RESs [30]. Since the RESs are usually islanded systems or located in weak ac grids, the HVDC tapping converter should employ grid-forming (GFM) control. Additionally, these RESs near the LCC-HVDC corridor typically have a modest scale, ranging from tens to hundreds of MW, compared to the GW scale of the LCC-HVDC corridor. As a result, the parallel tapping schemes [31]-[32], commonly used for tapping higher power levels (more than 20% of the LCC-HVDC system rating) [33]-[34], may face economic challenges due to the connection to the full HVDC voltage [29]. Although many DC-DC converters have been proposed in [29]-[30] and [35]-[37] to implement parallel tapping schemes at low power levels (less than 10% of the LCC-HVDC corridor), the costs remain high. Furthermore, further development of DC/DCs for HV to MV conversions is still required in engineering applications.

Based on the analyses above, the use of GFM series tapping converters to integrate relatively small-scale RESs to the LCC-HVDC corridor presents a promising solution for increasing the proportion of clean energy in the corridor, as shown in Fig.1. Additionally, the scheme in Fig.1 also can effectively mitigate the dc voltage drop issue in very long-distance HVDC systems. This is due to the dc-voltage-boosting effect provided by the HVDC series tapping converter integrated with RESs. That is to say, the scheme of using GFM series tapping converters to integrate small RESs to the LCC-HVDC also offers a feasible networking method for achieving very long-distance HVDC transmission systems. Based on these two merits of Fig.1, it is evident that the GFM HVDC series tapping converter holds great potential for facilitating the utilization of clean energy and addressing the new challenges of HVDCs in the process of achieving China's dual-carbon goals.

*B. Research Status of GFM HVDC Series Tapping Converter*

Although the concept of HVDC series tapping converters has been proposed for decades, most solutions have focused on extracting power from HVDC to supply nearby loads [33], [38]-[39]. With the recent rapid development of RESs, some studies have started exploring the integration of RESs using HVDC series tapping stations. In [40], diode rectifiers (DRs) are used as HVDC series tapping converters to integrate offshore wind farms (OWFs) into LCC-HVDCs. However, since DRs lack GFM control capability, this solution requires changing the existing control of the offshore wind farm and turbines. In [41]-[43], the self-commutating current source converter (CSC) is employed as an HVDC series tap to achieve GFM control. Nevertheless, the selection of switches and feasible switch-series techniques still need to be addressed.

Indeed, by modifying the control method, FBMMCs can be transformed into HVDC series tapping converters with great performance [18], [44]. However, although the application of FBMMC in HVDCs is becoming mature [9]-[11], the cost and power losses of FBMMC remains relatively high. Meanwhile, the footprint and weight of FBMMCs pose challenges in certain scenarios, such as offshore RESs. Last but not least, when FBMMCs are connected in series with LCCs, the disparities in current-carrying capacity between thyristors and full-controlled switches require the use of parallel technology for FBMMCs. This parallel technology includes switch-level, arm-level, and converter-level parallels. For instance, in Baihetan project [11], one LCC is connected in series with three parallel MMCs to match the current-carrying capacity. As a result, coordinating the operation of the parallel MMCs is highly complex.

To address the limitations mentioned above regarding the use of FBMMCs as GFM series tapping converters, the hybrid voltage source converter (VSC) topologies based on LCC and MMC provide inspiration. Consequently, this paper proposes a novel GFM HVDC series tapping converter, called Extended Flex-LCC (EFLCC), which is based on the extended techniques of the Flexible Line-Commutated Converter (Flex-LCC) [1]. To the best of the authors' knowledge, extensive analyses [25]-[27] have been conducted on hybrid topologies using LCC and MMC. However, these investigations mainly focus on using hybrid topologies to achieve VSCs [25]-[27], [45]-[46], as discussed in [1] regarding the Flex-LCC. Currently, there is no research specifically focused on investigating the feasibility of employing hybrid structures in the GFM HVDC series tapping converter. This paper aims to fill this research gap and also investigates the merits of the proposed EFLCC. The subsequent sections of this paper provide a detailed analysis.

## II. STEADY-STATE ANALYSIS OF THE INVESTIGATED SYSTEM

As depicted in Fig. 1, the system investigated in this paper comprises a primary LCC-HVDC corridor connected in series with several auxiliary RESs through HVDC tapping converters. The RESs in Fig. 1 are located near the LCC-HVDC, thereby, eliminating the need for long-distance networking. Moreover, the capacity of these RESs is relatively small compared to the capacity of the LCC-HVDC corridor, typically constituting less than 10% of its capacity. Furthermore, due to the corridor's extensive coverage across multiple RES-rich regions, different RESs can be distributed across various areas.

The equivalent circuit of the investigated system in the steady state is presented as Fig.2. In Fig. 2, $U_{dc0,R}$ and $U_{dc0,I}$ represent the ideal no-load dc voltages of the LCC rectifier and LCC inverter, respectively; $\alpha_R$ is the firing angle of the LCC rectifier, and $\gamma_I$ is the extinction angle of the LCC inverter, respectively; $R_{LCCR}$ and $R_{LCCI}$ denote the equivalent commutation resistances of the LCC rectifier and LCC inverter, respectively; $U_{dcR}$ and $U_{dcI}$ are the dc-port voltages of the LCC rectifier and LCC inverter, respectively; $I_{dc}$ is the dc current; $R_{line}$ is the equivalent resistance of the dc transmission lines; $\Delta U_{line}$ is the total voltage drop across the dc transmission lines; $\Delta U_{tap\Sigma}$ is the total voltage boost provided by all HVDC tapping converters.

Meanwhile, the following equations are true [47]:

$$\begin{cases} U_{dcR} = U_{dc0,R}\cos(\alpha_R) - I_{dc}R_{LCCR} \\ U_{dcI} = U_{dc0,I}\cos(\gamma_I) + I_{dc}R_{LCCI} \end{cases} \tag{1}$$

where

$$\begin{cases} U_{dc0,R} = \dfrac{6\sqrt{3}}{\pi}U_{ac,R} \\ U_{dc0,I} = \dfrac{6\sqrt{3}}{\pi}U_{ac,I} \end{cases} \text{ and } \begin{cases} R_{LCCR} = \dfrac{6}{\pi}X_{LCCR} \\ R_{LCCI} = \dfrac{6}{\pi}X_{LCCI} \end{cases} \tag{2}$$

where $U_{ac,R}$ and $U_{ac,I}$ are the line-to-neutral peak magnitude voltages of the rectifier and inverter valve sides, respectively; $X_{LCCR}$ and $X_{LCCI}$ are the commutation reactances, respectively.

Besides, the steady-state equations relating $U_{dcR}$, $U_{dcI}$, $\Delta U_{line}$ and $\Delta U_{tap\Sigma}$ can be derived as:

$$U_{dcR} - U_{dcI} = \Delta U_{line} - \Delta U_{tap\Sigma} \tag{3}$$

where

$$\begin{cases} \Delta U_{line} = I_{dc}R_{line} \\ \Delta U_{tap\Sigma} = \sum_{i=1}^{m}\Delta U_{tapi} = \dfrac{\sum_{i=1}^{m}P_{tapi}}{I_{dc}} = \dfrac{P_{tap\Sigma}}{I_{dc}} \end{cases} \tag{4}$$

where $\Delta U_{tapi}$ is the dc voltage boost provided by HVDC tapping converter $i$; $P_{tapi}$ is the active power provided by HVDC tapping converter $i$; $P_{tap\Sigma}$ is the total active power provided by HVDC tapping converters.

According to (1) ~ (4), when $I_{dc}$ and $U_{dcI}$ of the LCC-HVDC corridor are actively controlled by the rectifier and the inverter respectively, the total voltage drop $\Delta U_{line}$ can be compensated by the dc voltage boost provided by the tapping converters connected to RESs. As a result, the dc voltage distribution of the whole HVDC system can be optimized, addressing the two challenges described in the Introduction.

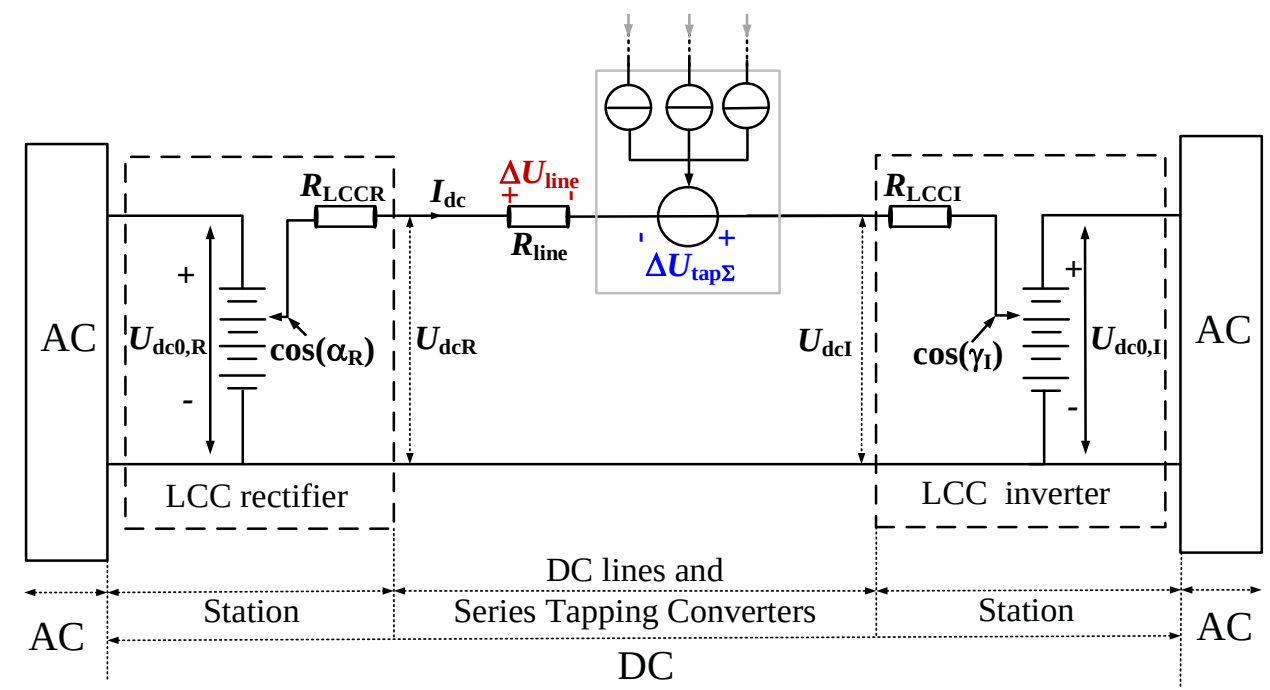

**Fig. 2.** Equivalent circuit of the investigated system in the steady state.

When implementing the solution of Fig. 1 to address the first challenge mentioned in the Introduction, since the parameters of the already built LCC-HVDC corridor cannot be changed, the maximum transmission proportion of clean energy in the reconstructed HVDC system can be expressed as:

$$\eta_{max} = \frac{P_{LCCR,RESs} + P_{tap\Sigma,max}}{P_{LCC\text{-}HVDC,ra}} \tag{5}$$

where $P_{LCC\text{-}HVDC,ra}$ is the rated active power of the LCC-HVDC corridor; $P_{LCCR,RESs}$ is the total active power of RESs allowed by the sending-end ac grid of the corridor; $P_{tap\Sigma,max}$ is the maximum value of $P_{tap\Sigma}$ allowed by the corridor, which is:

$$\begin{cases} \dfrac{P_{tap\Sigma,max}}{I_{dc,b}} - I_{dc,b}R_{line} = U_{dcI,max} - U_{dcR,min} \\ I_{dc,b} = \dfrac{P_{LCC\text{-}HVDC,ra} - P_{tap\Sigma,max}}{U_{dcR,min}} \qquad R_{line} = 2R_0L_{line} \end{cases} \tag{6}$$

where $U_{dcR,min}$ is the minimum dc-port voltage allowed by the LCC rectifier; $U_{dcI,max}$ is the maximum dc-port voltage allowed by the LCC inverter; $I_{dc,b}$ is the dc current boundary value; $R_0$ is resistance per unit length of each dc transmission line; $L_{line}$ is the length of each dc transmission line.

When the second challenge mentioned in the Introduction is adressed by Fig. 1, the rating design of the dc system shall satisfy the following steady-state equations:

$$U_{dcR,ra} - U_{dcI,ra} = I_{dc.ra}R_{line} - \frac{P_{tap\Sigma,ra}}{I_{dc,ra}} \tag{7}$$

$$P_{R,ra} + P_{tap\Sigma,ra} = P_{I,ra} \tag{8}$$

where $U_{dcR,ra}$ is the rated dc-port voltage of the LCC rectifier; $U_{dcI,ra}$ is the rated dc-port voltage of the LCC inverter; $I_{dc,ra}$ is the rated dc current of the investigetd system; $P_{tap\Sigma,ra}$ is the rated value of $P_{tap\Sigma}$ designed in the investigetd system; $P_{R,ra}$ is the rated active power of the LCC rectifier; $P_{R,ra}$ is the rated active power of the LCC inverter.

To gain a more intuitive understanding of the application effects and characteristics of (5) and (7), Fig. 3 presents the analysis results of the application effects within a ±800 kV LCC-HVDC system for the scheme depicted in Fig.1. In Fig. 3(a), the parameters of the existing LCC-HVDC are assumed to be as follows: $R_0$ = 6/1000 Ω/km [18]; $P_{\text{LCC-HVDC,ra}}$= 8000 MW; $P_{\text{LCCR,RESs}}$= 1500 MW; $L_{\text{line}}$=2000 km; $U_{\text{dcR,min}}$=($k_1$*±800) kV; and $U_{\text{dcI,max}}$ = ($k_2$*±800) kV. Then, according to (5) and (6), the variation of $\eta_{\max}$ with $k_1$ and $k_2$ is shown in Fig. 3(a). As can be seen from Fig. 3(a), the scheme in Fig. 1 can significantly increase the proportion of clean energy in the LCC-HVDC corridor without reducing the strength of the sending-end ac grid. Additionally, the greater the difference between $U_{\text{dcR,min}}$ and $U_{\text{dcI,max}}$, the more clean energy can be transmitted by the proposed soluation. In particular, when $U_{\text{dcR,min}}$=(0.9*±800) kV, $U_{\text{dcI,max}}$ = (1.1*±800) kV, the scheme used in Fig. 1 increases the transmission ratio of clean energy in the LCC-HVDC from 19% to 41%, validating the application effect of (5). In Fig. 3(b), the parameters of the dc system are assumed to be as follows: $R_0$ = 6/1000 Ω/km [18]; $U_{\text{dcR,ra}}$=($k_3$*±800) kV; $U_{\text{dcI,ra}}$ =±800kV; $I_{\text{dc,ra}}$= 5kA. Then, the variation of $L_{\text{line}}$ with $P_{\text{tap}\Sigma\text{,ra}}$ is shown in Fig. 3(b). When $P_{\text{tap}\Sigma\text{,ra}}$=0, the transmission distance of the original scheme increases with an increase in $k_3$. For instance, when $k_3$=1.1, the maximum transmission distance of the original scheme is 2600km. Meanwhile, the proposed solution significantly enhances the transmission distance of the dc system. For example, when $k_3$=1.1 and $P_{\text{tap}\Sigma\text{,ra}}$ =400MW (5% of the rated capacity of the LCC-HVDC corridor), the maximum transmission distance of the dc system increases to 4000km. Similarly, when $k_3$=1.1 and $P_{\text{tap}\Sigma\text{,ra}}$ =800MW (10% of the rated capacity of the LCC-HVDC), the maximum transmission distance of the dc system increases to 5300km.

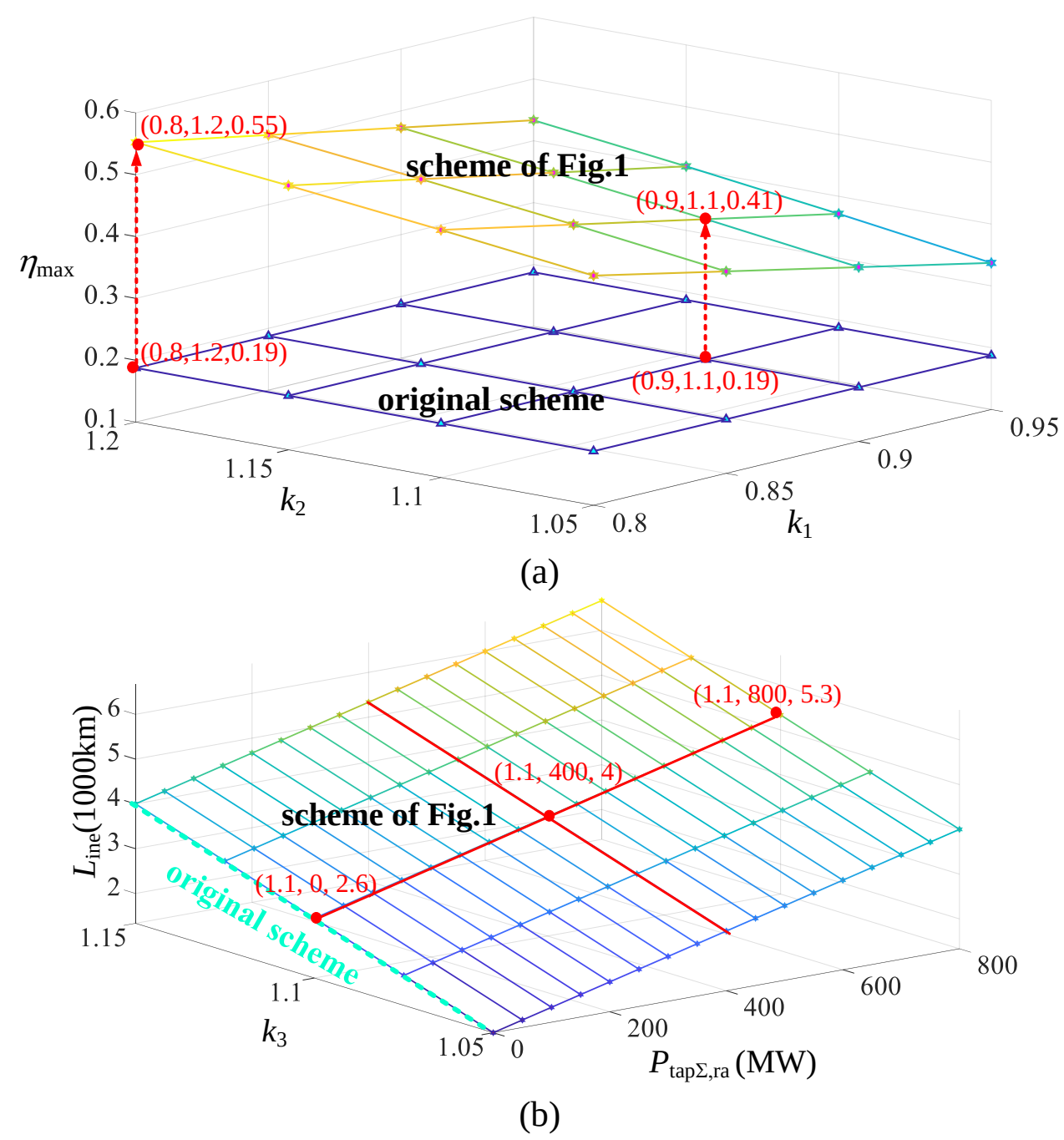

**Fig. 3.** The application analysis of the soluation in Fig. 1. (a) application effect of (5), (b) application effect of (7)

## III. Implementation of The Proposed EFLCC

Section II analyzes and demonstrates the advantages of the investigated solution presented in Fig. 1, focusing on its ability to enhance the utilization rate of clean energy and increase the maximum transmission distance of the LCC-HVDC corridor. It is evident that the HVDC tapping converters play a crucial role in Fig. 1. Therefore, in order to promote the implementation of Fig. 1, a novel techno-economic GFM HVDC series tapping converter, called EFLCC, is proposed in this section.

Fig. 4 presents the main circuit and control scheme of the proposed EFLCC. Similar to the Flex-LCC [1], the EFLCC also utilizes a hybrid-valve topology based on LCC and FBMMC and uses the control framework proposed in [1]. Specifically, the EFLCC maintains the series-connected configuration of LCC and FBMMC. Meanwhile, the firing angle of the LCC provides no control variable for the EFLCC, while the FBMMC allows for three control variables. However, the specific control scheme and parameter design methodology in [1] are proposed to operate the hybrid valve as a VSC. Therefore, it is necessary to extend these techniques to operate the hybrid valve shown in Fig. 4 (a) as a GFM HVDC series tap, i.e., to achieve the EFLCC. The following sections provide an in-depth analysis.

### *A. Modified Arm-Internal-Electromotive-Force Control*

Fig. 4(b) presents the detailed control scheme to achieve the proposed EFLCC. In Fig. 4(b), "*" and "m" are the reference and measure values, $\alpha$ is the firing angle of the LCC part, "d" and "q" are the values of the $d,q$ sequences; $V_{\text{gd}}$ and $V_{\text{gq}}$ are the $d,q$ components of $v_{\text{ga,b,c}}$, $V_{\text{cap}\Sigma}$ is the average value of the sum of capacitor voltages per arm in the FBMMC part, $i'_{\text{Ma,b,c}}$ is the grid-side currents of the FBMMC part, and $i'_{\text{Md}}$ and $i'_{\text{Mq}}$ are the $d,q$ components of $i'_{\text{Ma,b,c}}$, $k_{\text{M}}$, $L_{\text{cal}}$, $u_{\text{Ma,b,c}}$ and $E_{\text{dcM}}$ are the transformer turn, equivalent inductance, valve-side ac voltages and arm-internal-electromotive-force voltage of the FBMMC part. The subscripts "p" and "n" are the upper and lower arms of the FBMMC part.

The six arm voltages $u_{\text{a,b,cp}} \sim u_{\text{a,b,cn}}$ of the FBMMC part in the EFLCC can be derived as:

$$\begin{cases} u_{\text{a,b,cp}} = \dfrac{E_{\text{dcM}}}{2} - u_{\text{Ma,b,c}} \\ u_{\text{a,b,cn}} = \dfrac{E_{\text{dcM}}}{2} + u_{\text{Ma,b,c}} \end{cases} \tag{9}$$

The equivalent inductance $L_{\text{cal}}$ in Fig. 4(b) is:

$$L_{\text{cal}} = k_{\text{M}} L_{\text{eM}} + \frac{L_{\text{arm}}}{2k_{\text{M}}} \tag{10}$$

where $L_{\text{eM}}$ is the equivalent inductance of the transformer in the FBMMC part, and $L_{\text{arm}}$ is the equivalent inductance of the arm inductor in the FBMMC part.

It can be observed in Fig. 4(b) that the control framework of the EFLCC is similar to that of the Flex-LCC [1], with the main distinction lying in the control of arm internal electromotive force. That is, the generation method for the reference values of $E_{\text{dcM}}$ in the EFLCC and the Flex-LCC differs. This is due to the different application conditions for the EFLCC and the Flex-LCC, as shown in Fig. 5. In Fig. 5(a), the external dc system controls the dc current of the whole dc system, while in Fig. 5(b), it controls the dc voltage. As a result, the characteristics of the dc-port current in the EFLCC will significantly differ from its characteristics in the Flex-LCC, implying that the dc-port current of the EFLCC will not exhibit the same dynamics with that of the Flex-LCC. Therefore, the dynamic equation for $E_{\text{dcM}}$ in the EFLCC will be transformed from equation (11) (the Flex-LCC) to equation (12):

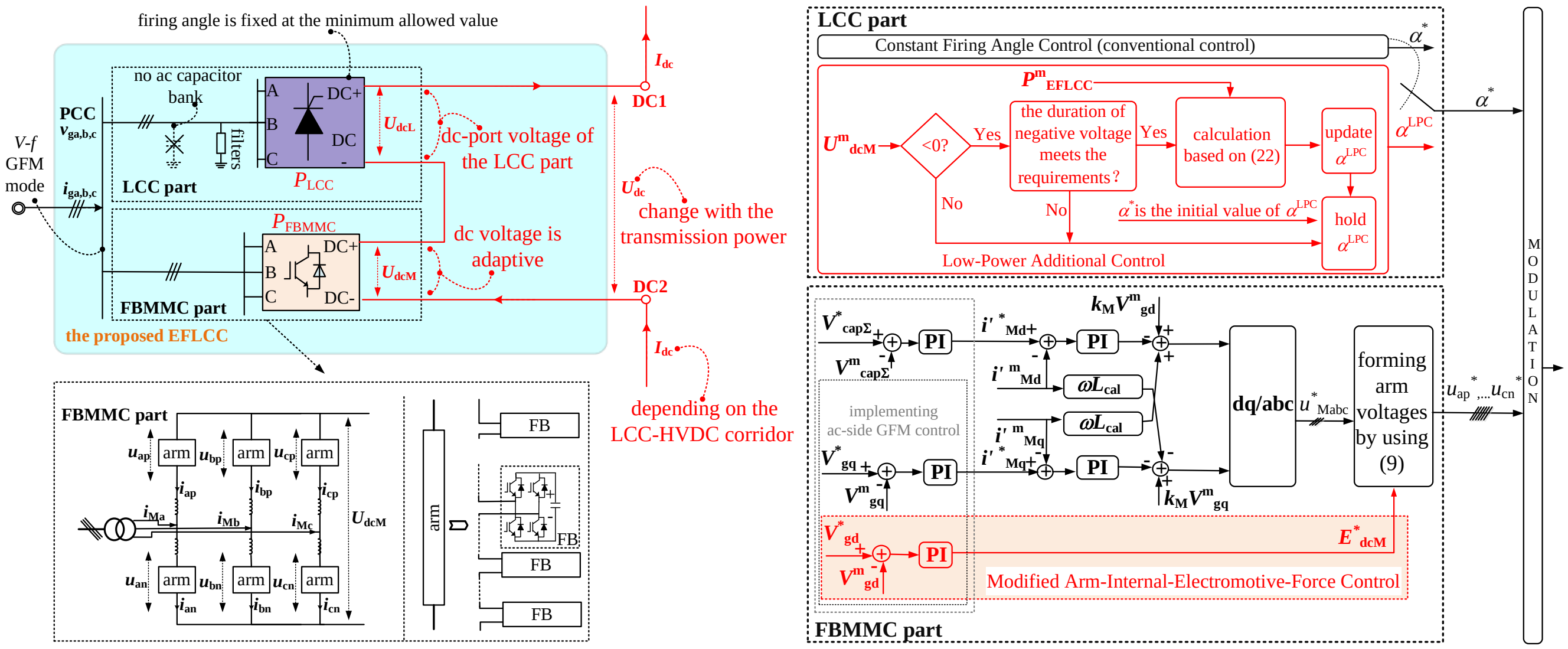

**Fig. 4.** The main circuit and control scheme of the proposed EFLCC (the red words and lines highlight the main differences between the proposed EFLCC and the Flex-LCC [1]). (a) main circuit , (b) control scheme.

$$E_{dcM} = \frac{L_{arm}}{6}\frac{dI_{dc}}{dt} + \frac{U_{dcM}}{2} \quad (11)$$

$$E_{dcM} = \frac{U_{dcM}}{2} \quad (12)$$

Because the EFLCC operates in the GFM mode, $V_{gd}$ directly affects the dynamics of $U_{dcM}$. This allows the arm-internal-electromotive-force control loop of the EFLCC to be designed based on (12), as shown in Fig. 4(b). As a result, Fig. 4(b) enables the EFLCC to work as a GFM HVDC series tapping converter, as shown in Fig. 5(a), rather than a GFM VSC.

*B. Converter-level Operation Analysis for the EFLCC*

Comparing Figs. 5(a) and 5(b), the operation of the EFLCC are distinct from those of the Flex-LCC due to the modified control method in Fig. 4(b).

When the EFLCC operates in a steady state, the following equations hold true:

$$U_{dcL} = \frac{6\sqrt{3}}{\pi}k_L V_g \cos\alpha^* - \frac{6}{\pi}k_L{}^2\omega_g L_{eL} I_{dc} \quad (13)$$

and

$$\begin{cases} U_{dc} = \dfrac{P_{EFLCC}}{I_{dc}} \\ U_{dcM} = U_{dc} - U_{dcL} \end{cases} \quad (14)$$

where $k_L$ is the transformer turns of the LCC part in the EFLCC; $L_{eL}$ is the equivalent inductance of the transformer in the LCC part; $\omega_g$ is the angular frequency of $v_{ga,b,c}$, $P_{EFLCC}$ is the active power value of the EFLCC; $V_g$ is the voltage amplitude of $v_{ga,b,c}$, and can be derived as:

$$V_g = \sqrt{V_{gd}^2 + V_{gq}^2} \quad (15)$$

According to (13) and (14), when the EFLCC operates in a steady state with fixed $I_{dc}$, $V_g$, and $\omega_g$, $U_{dcL}$ remains constant regardless of the transmission power due to the fixed value of $\alpha^*$. This means that when the external system controls $I_{dc}$, the transmission power of the LCC part of the EFLCC remains unchanged because $U_{dcL}$ and $I_{dc}$ both remain unchanged. In this scenario, the transmission power change of the EFLCC depends on the $U_{dc}$ change caused by the $U_{dcM}$ change. That is, when $I_{dc}$ is fixed, the power change in the EFLCC is solely determined by the power change in its internal FBMMC part and is not affected by the LCC part. These operational characteristics distinguish the EFLCC from existing hybrid structures of GFM VSC, such as the Flex-LCC [1] and others [25]-[27], [45]-[46], resulting in different operating boundaries for the main circuit. The detailed analysis is as follows.

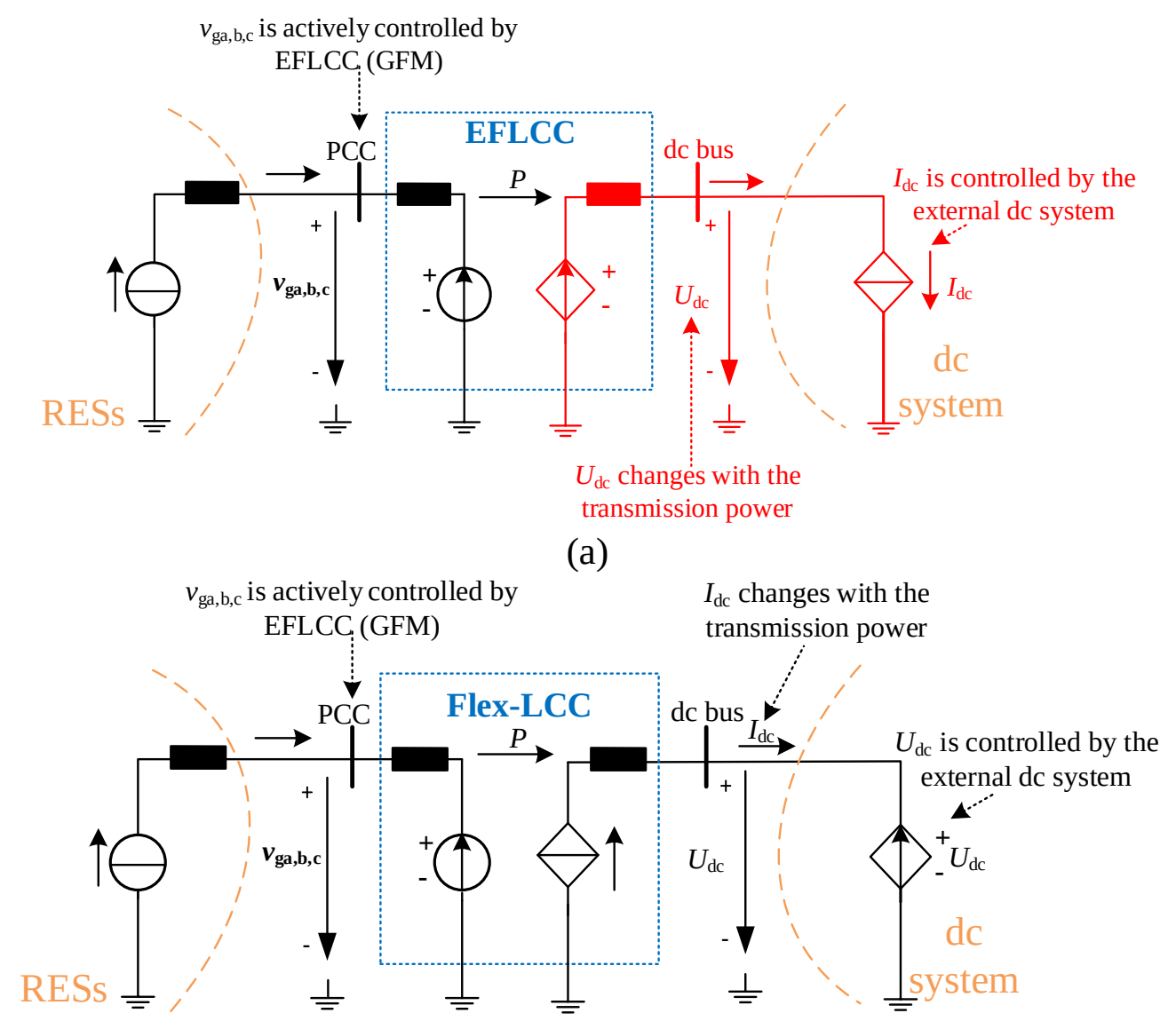

(b)

**Fig. 5.** Macroscopic system-level models of the EFLCC and the Flex-LCC. (a) the EFLCC, (b) the Flex-LCC.

Because the reactive power of the LCC part in the EFLCC is determined by its active power transmission value and $\alpha^*$. The following equations can be derived based on (13):

$$\begin{cases} P_{LCC_\alpha^*} = \dfrac{6\sqrt{3}}{\pi}k_L V_g \cos\alpha^* I_{dc} - \dfrac{6}{\pi}k_L{}^2\omega_g L_{eL} I_{dc}^2 \\ Q_{LCC_\alpha^*} = P_{LCC_\alpha^*}\sqrt{\left(\dfrac{\sqrt{3}V_g}{\sqrt{3}V_g\cos\alpha^* - k_L\omega_g L_{eL} I_{dc}}\right)^2 - 1} \end{cases} \quad (16)$$

where $P_{LCC_\alpha*}$ and $Q_{LCC_\alpha*}$ are the active and reactive power values of the LCC part, respectively, when the EFLCC operates at $I_{dc}$, $P_{EFLCC}$, $Q_{EFLCC}$, and the firing angle of the LCC part is $\alpha^*$. Meanwhile, $Q_{EFLCC}$ is the reactive power value of the EFLCC.

Obviously, (16) shows that when the EFLCC operates in a specific steady state, the reactive power of the LCC part is similar to its active power and remains constant regardless of the transmission power of the EFLCC due to the fixed value of $\alpha^*$. As a result, the apparent power of the LCC part $S_{LCC_\alpha*}$ also remains constant regardless of the transmission power of the EFLCC because $S_{LCC_\alpha*}$ can be derived as follows:

$$S_{LCC_\alpha^*} = \sqrt{\left(P_{LCC_\alpha^*}\right)^2 + \left(Q_{LCC_\alpha^*}\right)^2} \tag{17}$$

Then, according to Fig. 4(a), in the EFLCC, both $Q_{LCC}$ and $Q_{EFLCC}$ are compensated by the FBMMC part (ignoring the reactive power compensated by the ac filters). Therefore, the following equations can be derived based on (14):

$$\begin{cases} P_{FBMMC} = P_{EFLCC} - P_{LCC_\alpha^*} \\ Q_{FBMMC} = Q_{LCC_\alpha^*} + Q_{EFLCC} \end{cases} \tag{18}$$

and

$$S_{FBMMC} = \sqrt{\left(P_{FBMMC}\right)^2 + \left(Q_{FBMMC}\right)^2} \tag{19}$$

where the positive direction of $Q_{LCC_\alpha*}$ indicates reactive power flows from the PCC to the LCC part; the positive direction of $Q_{FBMMC}$ indicates that reactive power flows from the FBMMC part to the PCC; the positive direction of $Q_{EFLCC}$ indicates that reactive power flows from the EFLCC to the PCC.

Based on (18) - (19), as $P_{LCC_\alpha*}$ and $Q_{LCC_\alpha*}$ remain constant when $P_{EFLCC}$ and $Q_{EFLCC}$ change, the changes of $P_{EFLCC}$ and $Q_{EFLCC}$ will be directly reflected in $P_{FBMMC}$ and $Q_{FBMMC}$. Then, the simplified internal equivalent model of the EFLCC is shown as Fig. 6(a). Based on Fig. 6(a), it can be concluded that when the external operating conditions remain constant, i.e. when the values of $I_{dc}$ and $V_g$ are constant, the active and reactive power of the LCC part in the EFLCC depend solely on the value of $\alpha^*$ and are independent of the power transmitted by the EFLCC. Additionally, since the EFLCC follows the control principle similar to the Flex-LCC, the value of $\alpha^*$ is decoupled from the control system of the EFLCC. This implies that, if necessary, the operating performance of the EFLCC can be optimized by adjusting the value of $\alpha^*$ under certain conditions.

### *C. Discussion of Additional Coordination Strategy in Low-Power Operation*

According to the equivalent model shown in Fig. 6(a), the internal power distribution of the EFLCC can be illustrated in Fig. 6(b) as the power changes. It is evident that when $P_{EFLCC} > P_{LCC_\alpha*}$, the transmission power direction of $P_{FBMMC}$ is the same as $P_{EFLCC}$ and $P_{LCC_\alpha*}$. However, when $P_{EFLCC} \leq P_{LCC_\alpha*}$, the transmission power direction of $P_{FBMMC}$ will be opposite to $P_{EFLCC}$ and $P_{LCC_\alpha*}$. In other words, in this case, the FBMMC part of the EFLCC exhibits power circulation, increasing the power losses of the tap converter. Thus, to avoid the occurrence of power circulation in the FBMMC part during low-power operation of the EFLCC, an additional low-power coordination control strategy (LPC) is proposed in this section to enhance the low-power operating efficiency of the EFLCC.

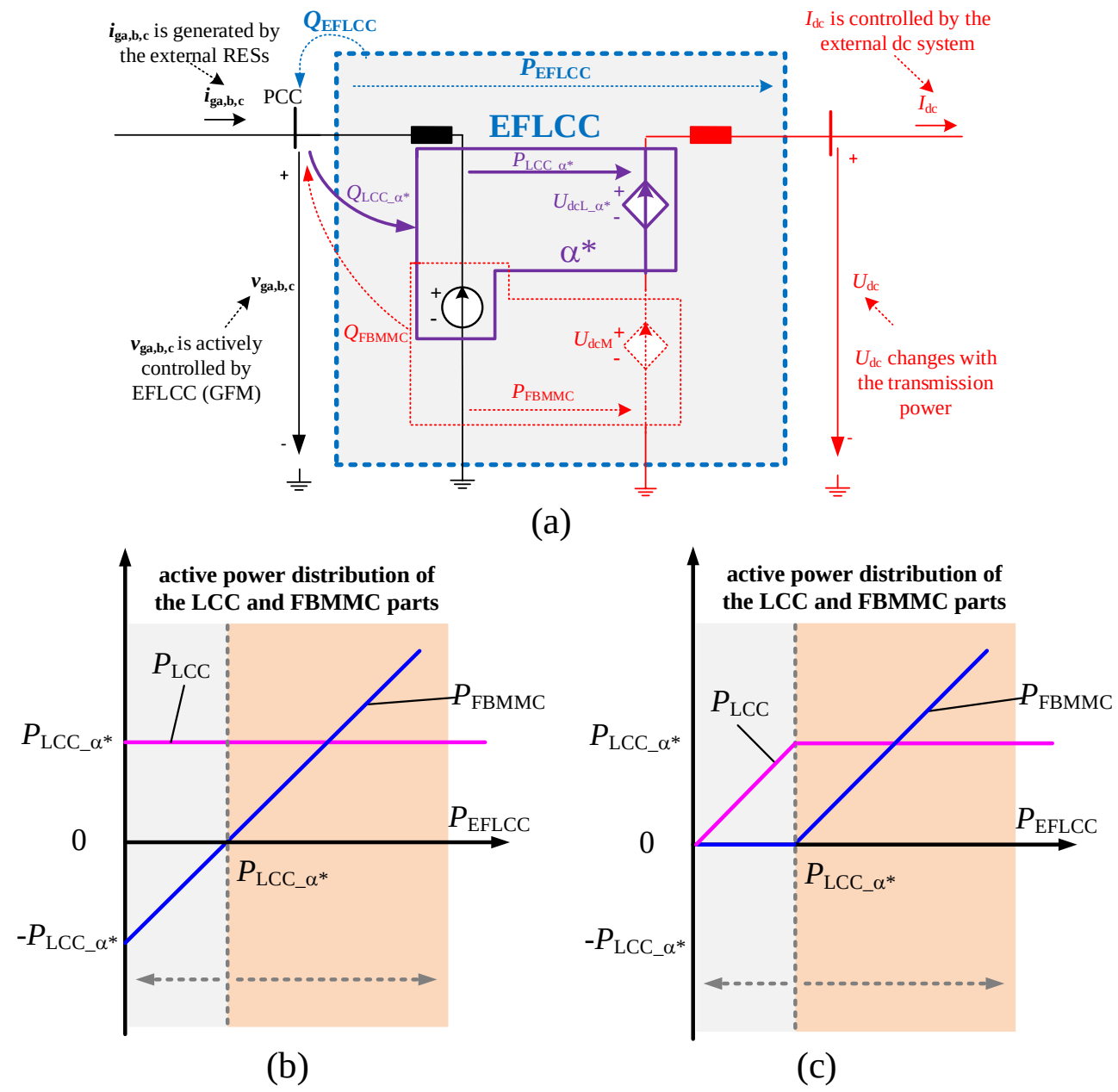

**Fig. 6.** The operating equivalent model and power distribution analysis of the EFLCC. (a) equivalent model (b) no the LPC (c) with the LPC

According to Fig. 6(a) and 6(b), when the EFLCC operates in a low-power condition, the power circulation in the FBMMC part can be actively controlled by adjusting the value of $\alpha^*$. Considering that the LCC part has higher efficiency compared to the FBMMC part at the same transmission power, the main purpose of the LPC is to ensure that the dc-port voltage of the FBMMC part operates near 0 V when the EFLCC operates at the low-power condition, i.e. when $P_{EFLCC} \leq P_{LCC_\alpha*}$. Then,

$$\begin{cases} P_{LCC_LPC} = \dfrac{6\sqrt{3}}{\pi} k_L V_g \cos\alpha^{LPC} I_{dc} - \dfrac{6}{\pi} k_L{}^2 \omega_g L_{eL} I_{dc}^2 = P_{EFLCC} \\ Q_{LCC_LPC} = P_{EFLCC} \sqrt{\left( \dfrac{\sqrt{3}V_g}{\sqrt{3}V_g \cos\alpha^{LPC} - k_L \omega_g L_{eL} I_{dc}} \right)^2 - 1} \end{cases} \tag{20}$$

and

$$\begin{cases} P_{FBMMC_LPC} = P_{EFLCC} - P_{LCC_LPC} = 0 \\ Q_{FBMMC_LPC} = Q_{LCC_LPC} + Q_{EFLCC} \end{cases} \tag{21}$$

where $P_{LCC_LPC}$ and $Q_{LCC_LPC}$ are the active and reactive power values of the LCC part, respectively, when the EFLCC operates at the low-power conditions; $P_{FBMMC_LPC}$ and $Q_{FBMMC_LPC}$ are the active and reactive power of the FBMMC part, respectively, when the EFLCC operates at the low-power conditions; $\alpha^{LPC}$ is the firing angle generated by the LPC, and can be derived as:

$$\alpha^{LPC} = \mathrm{a}\cos\left( \frac{\pi}{6\sqrt{3}} \frac{P_{EFLCC}}{k_L V_g I_{dc}} + \frac{k_L \omega_g L_{eL} I_{dc}}{\sqrt{3} V_g} \right) \tag{22}$$

As shown in (20)-(22), when the EFLCC operates under the low-power conditions, the lower the active power transmitted by the EFLCC, the larger the value of $\alpha^{LPC}$ should be in order to satisfy the dc-port voltage change of the EFLCC. In Addition, when the LPC is utilized, the internal power distribution of the EFLCC can be depicted as shown in Fig. 6(c). Comparing with Fig. 6(b), when the EFLCC operates in low power, i.e. $P_{EFLCC} \le P_{LCC_\alpha*}$, the active power of the LCC part does not remain at $P_{LCC_\alpha*}$, but decreases as $P_{EFLCC}$ decreases. As a result, the reactive power of the LCC part also varies with $P_{EFLCC}$. It should be noted that when the power of the EFLCC, $P_{EFLCC}$, is extremely small, the firing angle, $\alpha^{LPC}$, of the LCC part will be very large, resulting in a large reactive power that needs to be absorbed by the LCC part, as shown in (20) and Fig. 6(c). This increase the reactive capacity of the FBMMC part, and should be a primary consideration in the design.

In conclusion, when the LPC is used in low-power conditions, the power circulation of the FBMMC part can be eliminated, resulting in an improved average operating efficiency of the EFLCC. However, the control and design method will be more complex compared to the EFLCC without the LPC.

## V. Main-Circuit Parameters Design and Comparison

### *A. Design Analysis of the Transformer Turns Ratio for the LCC Part*

In the EFLCC, the different capacity design ratios of the LCC and FBMMC parts are closely related to the cost. From the perspective of active power capacity design, the larger the active power capacity of the LCC part in the EFLCC, the more beneficial it is to reduce the cost due to the cost per unit of active power capacity of the FBMMC part is higher than that of the LCC part. However, according to the analysis in Section III.B, in the EFLCC, both the reactive power capacity absorbed by the LCC part and the reactive power capacity generated by the whole EFLCC need to be supplemented by the FBMMC part. In this way, from the perspective of reactive power capacity design, the smaller the reactive power capacity of the LCC part, the more beneficial it is to reduce the cost of the converter due to the reduction of reactive power capacity of the FBMMC part. In fact, since the reactive power capacity absorbed by the LCC component is closely related to its active power capacity [47]-[48], in the parameter design of the EFLCC, we can target the minimum apparent capacity of the FBMMC component to select the active power capacity of the LCC component, thereby improving the economy of the EFLCC.

Assuming the rated conditions of the EFLCC, such as $<V_g>$, $<\omega_g>$, $<I_{dc}>$, $<P_{EFLCC}>$ and $<Q_{EFLCC}>$ , are known, and $L_{eL}$ is designed based on [47]. The transformer ratio of the LCC part, $k_L$, should be designed as follows.

When the LPC is not used, given that the EFLCC is used to connect to RESs, it is essential to ensure that the power operating range extends from 0 to the rated value, regardless of the infrequent occurrence of low power operating conditions due to meticulous planning of the RES power plants, this range still need to be maintained. Then, $<k_{L_\alpha*}>$ should satisfy:

$$\begin{cases} \left\langle P_{LCC_\alpha^*} \right\rangle \le \dfrac{\left\langle P_{EFLCC} \right\rangle}{2} \\ \min\left[\left\langle S_{FBMMC} \right\rangle\right] \end{cases} \tag{23}$$

where $<k_{L_\alpha*}>$ is the designed value of $k_L$ ; min[] represents the minimum value function.

When the additional LPC is used in the EFLCC, it is essential to make the design value of $k_L$ effective for both the scenario where $\alpha^*$ remains fixed and the scenario where the additional LPC is used. Thus, $<k_{L_LPC}>$ should satisfy:

$$\begin{cases} \dfrac{\left\langle P_{EFLCC} \right\rangle}{2} \le \left\langle P_{LCC_\alpha^*} \right\rangle \le \left\langle P_{EFLCC} \right\rangle \\ \min\left[\left\langle S_{FBMMC} \right\rangle\right] \end{cases} \tag{24}$$

where $<k_{L_LPC}>$ is the designed value of $k_L$ when the additional LPC is used in the EFLCC.

In fact, given the operating characteristics of the LCC part shown in (16)~(17) and (20)~(21), regardless of whether the LPC is used, the apparent capacity of the FBMMC part is close to its minimum value at $<P_{LCC_\alpha*}> = 1/2 <P_{EFLCC}>$. Thus, the design value of $<k_{L_\alpha*}>$ can be easily and directly obtained from:

$$\frac{6\sqrt{3}}{\pi}\left\langle V_g \right\rangle \cos\alpha^* k_L - \frac{6}{\pi}\left\langle \omega_g \right\rangle \left\langle L_{eL} \right\rangle \left\langle I_{dc} \right\rangle k_L^2 = \frac{\left\langle P_{EFLCC} \right\rangle}{2\left\langle I_{dc} \right\rangle} \tag{25}$$

Based on (25), the designed value of $k_L$ can be determined. Then, the designed values of the active power capacity, reactive power capacity, and overall capacity for both the LCC and FBMMC parts can be calculated based on (16)-(19) with $<V_g>$, $<\omega_g>$, $<I_{dc}>$, $<P_{EFLCC}>$, $<Q_{EFLCC}>$, $<L_{eL}>$, $<k_L>$ and $<\alpha^*>$.

### *B. Design Analysis of the Current-carrying Capacity for the FBMMC Part*

In the proposed EFLCC, the FBSMs within the FBMMC part can choose between charge and discharge operation modes, regardless of the current direction in each arm [49]. Therefore, once the capacity design value of the FBMMC part, $<S_{FBMMC}>$, is determined based on Section V.A, the current-carrying capacity of the full-controlled switches can be optimized by using the transformer turns ratio of the FBMMC part, $k_M$.

For the FBMMC part of the EFLCC, the root mean square (RMS) value and maximum value of arm currents are:

$$\begin{cases} \mathrm{RMS}_{i_{FBMMC}} = \dfrac{1}{3}\sqrt{\left\langle I_{dc} \right\rangle^2 + \dfrac{1}{2}\left(\dfrac{\left\langle S_{FBMMC} \right\rangle}{k_M\left\langle V_g \right\rangle}\right)^2} \\ \mathrm{MAX}_{i_{FBMMC}} = \dfrac{1}{3}\left(\left\langle I_{dc} \right\rangle + \dfrac{\left\langle S_{FBMMC} \right\rangle}{k_M\left\langle V_g \right\rangle}\right) \end{cases} \tag{26}$$

where $<S_{FBMMC}>$ is the designed capacity of the FBMMC part.

Based on (26), the valve-side ac voltage of the FBMMC part, $k_M<V_g>$, has a significant impact on the current-carrying capacity of full-controlled switches when the capacity of the FBMMC part is determined. Then, to ensure the matching of the current-carrying capacities between full-controlled switches and thyristors, $k_M$ should be designed as:

$$\left\langle k_M \right\rangle \ge \frac{\left\langle S_{FBMMC} \right\rangle}{\left\langle V_g \right\rangle}\mathbf{max}\left[\frac{1}{\sqrt{2\left(\left(3\mathrm{RMS}_{sch}\right)^2 - \left\langle I_{dc} \right\rangle^2\right)}}, \frac{1}{3\mathrm{MAX}_{sch} - \left\langle I_{dc} \right\rangle}\right] \tag{27}$$

where $<k_M>$ is the designed value of $k_M$ ; $\mathrm{RMS}_{sch}$ and $\mathrm{MAX}_{sch}$ are the limitations of RMS and maximum values for the full-controlled switches used in the FBMMC part; and the output of **max** $[x,y]$ is the larger value of $x$ and $y$.

Eq. (27) shows that the voltage design for the valve-side ac-and dc- ports of the FBMMC part in the proposed EFLCC is no longer confined to the conventional 0.5 times. Instead, they are designed to facilitate matching the current-carrying capacity between the full-controlled switches and thyristors.

TABLE I MAIN PARAMETERS OF THE STAUDY CASE

| Parameters | Values |
|---|---|
| $<I_{dc}>$ | 5 kA |
| $\sqrt{3/2}<V_g>$ | 66 kV |
| $<\omega_g>$, $<\omega_g><L_{eL}>$ | 50 Hz, 3Ω |
| rated active power | 400 MW |
| rated reactive power | 50 MVar |
| $\alpha^*$ | 5º |

TABLE II KEY PARAMETERS COMPARISON

| Topologies | FBMMC | EFLCC |
|---|---|---|
| $<k_L>$ | / | 15 kV/66 kV |
| $< S_{LCC} >$ | / | 208 MVA |
| $< U_{dcL}>$ | / | 40kV |
| $<k_M>$ | 40 kV/66 kV | 30 kV/66 kV |
| $< S_{MMC} >$ | 403 MVA | 228 MVA |
| $\mathrm{RMS}_{i_{FBMMC}}$ | 3.35 kA | 2.75 kA |
| $\mathrm{MAX}_{i_{FBMMC}}$ | 5.78 kA | 4.7 kA |
| $<U_{dcM}>$ | 80kV | 40kV |
| $<V^*_{cap\Sigma}>$ | 80kV | 60kV |
| estimated power losses (rated condition) | 4.35 MW | 3.77 MW |

### *C. Comparison Analysis*

To more intuitively analyze the effectiveness of the proposed EFLCC, the parameters listed in Table I are used as a case to conduct a detailed comparative analysis between the EFLCC and the FBMMC HVDC tap.

Based on Table I and Section V.A, the key parameters of the EFLCC can be obtained as presented in Table II. Comparing the analysis results in Table II, regardless of whether the LPC is used in the EFLCC, the capacity demand for the FBMMC part in the proposed EFLCC is always smaller than that in the FBMMC-based HVDC tap for the same rated active power and reactive power requirement. This characteristic facilitates the reduction of cost, weight, and volume in constructing a GFM HVDC tap station, even when considering the addition of the LCC part in the proposed EFLCC. In addition, the current-carrying capacity of the full-controlled switches can be matched with the thyristors more easily in the EFLCC due to the reduced capacity demand and the proposed design method in (27) of the FBMMC part. For example, in Table II, the RMS and maximum values of the arm currents in the FBMMC-based GFM HVDC tap and the FBMMC part of the EFLCC differ. Clearly, in the EFLCC, the full-controlled switches require a smaller current-carrying capacity (RMS value: 3.35 kA reduces to 2.75 kA and RMS value: 5.78 kA reduces to 4.70 kA) to match the current-carrying capacity of thyristors. These demonstrate the EFLCC is able to flexibly address the mismatch in current-carrying capacity between its full-controlled switches and thyristors, showcasing good practical utility.

In fact, the capacity demand reduction of the FBMMC part in the EFLCC not only helps to reduce the cost, weight, and volume of the GFM HVDC tapping station but also enhances the efficiency performance, as the LCC has lower power losses than the FBMMC. Based on the data from [50], the average losses per LCC station and per FBMMC station are 0.63% and 1.08%, respectively. Therefore, the estimated rated power losses of the FBMMC-based GFM HVDC tap and the EFLCC can be shown in Table II. Clearly, the proposed EFLCC will have a higher rated efficiency.

## VI. VALIDATION

### *A. Experimental Results of the EFLCC*

To validate the effectiveness of the proposed control principle for the EFLCC, experimental results of a prototype are presented in Fig. 7, along with detailed parameters in Table III.

In Fig. 7(a), steady-state results of the EFLCC prototype under $I_{dc}$= 3 A, $P_{tap}$= 1.5 kW are shown. The RMS value of the ac line-to-line voltage of the prototype is actively controlled at 95 V, validating the ac GFM control capability of the proposed EFLCC. In addition, the dc voltages of both the LCC and FBMMC parts in the prototype are approximately 250 V, while the dc-port voltage of the EFLCC in this case is 500 V. Similarly, the results in Fig. 7(b), where the EFLCC prototype operates under $I_{dc}$= 3 A, $P_{tap}$= 1 kW, further demonstrate that the EFLCC can function as a GFM HVDC tap. In addition, when the ac-port voltage is kept constant by the GFM control, $U_{dcL}$ within the tap remains constant while $U_{dcM}$ adjusts according to the transmission power, resulting in the adjustment of the tap's dc-port voltage $U_{dc}$. This further confirms the correctness of the theoretical analysis. In Fig. 4(c), the dynamic results of the EFLCC prototype are shown. The experiment reveals that as the power of the EFLCC changes, $U_{dcL}$ remains steady at 250 V, while $U_{dcM}$ fluctuates between 250 V and 83 V, causing the overall $U_{dc}$ of the EFLCC prototype to vary from 500 V to 333 V, further validating the analysis.

It is important to note that the experiment aimed to validate that the control principle and method depicted in Fig. 4 could allow the EFLCC to function as a GFM HVDC tap. Therefore, the signal source for the firing angle of the LCC part did not affect the validity of the results. Despite manually setting the firing angle of the LCC at a fixed value (5º) during the experiment, the results remain applicable to both the EFLCC without the LPC and the EFLCC with the LPC, as the use of LPC solely impacts the signal source of the firing angle without changing the core control principles and implementation of the control method in Fig. 4.

### *B. Simulation Results of the EFLCC -based HVDC System*

The simulation results of the proposed EFLCC's application in an HVDC system are presented in Fig. 8. In the simulation model, the system topology is configured as shown in Fig. 1, and the dc current $I_{dc}$ is controlled by the LCC-HVDC system. Meanwhile, the detailed parameters of the EFLCC are shown in Tables I and II.

Before $t_1$, the power of the EFLCC is 400 MW. The frequency and RMS value of the ac-port voltage of the tap are actively regulated at 50 Hz and 66 kV, while the dc-port voltage $U_{dc}$ is maintained at 80 kV. Meanwhile, the dc voltages of both the LCC and FBMMC parts, i.e. $U_{dcL}$ and $U_{dcM}$, in the EFLCC are 40 kV. As a result, the power of the LCC part ($P_{LCC}$) and FBMMC part ($P_{MMC}$) are both 200 MW. These validate the effectiveness of implementing the EFLCC in HVDC systems.

TABLE III MAIN PARAMETERS OF EXPERIMENTS

| Parameters | Experiment |
|---|---|
| $I_{dc}$ | 3 A |
| $\sqrt{3/2}V_g$, $\omega_g$ | 95 V, 50 Hz |
| $U_{dc}$, $P_{tap.max}$ | 0~500 V, 1500 W |
| $\alpha^*$ ( manual setting ) | 5º |
| $k_L$, $k_M$ | 1/1, no transformer |
| maximum reactive power of ac filters | 0 Var (no ac filter) |
| $V^*_{cap\Sigma}$ | 240 V |

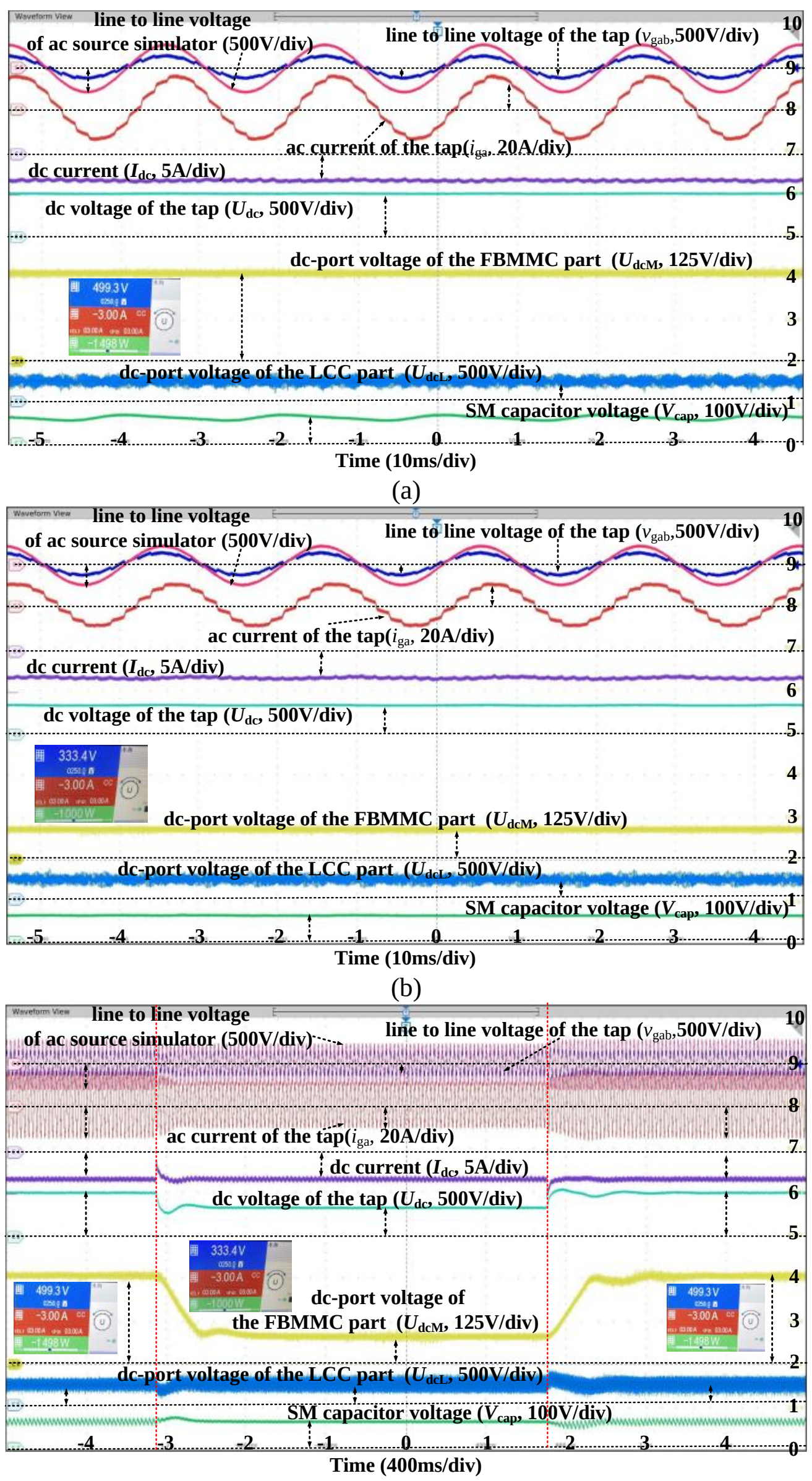

**Fig. 7.** Prototype experimental results of the proposed EFLCC. (a) steady-state results, 1500W, (b) steady-state results, 1000W, (c) dynamic results.

During $t_1$ to $t_2$, the transmission power of the EFLCC is 100 MW. The PLC is not used during $t_1$ to $t_{11}$, but it is utilized during $t_{11}$ to $t_2$. In this scenario, the variation in the operation of the EFLCC is mainly reflected in the internal distribution of dc voltage and power. Specifically, from $t_1$ to $t_{11}$, the dc-port voltage of the LCC part remains at 40 kV, while the dc-port voltage of the FBMMC part changes to -20 kV. As a result, the dc-port voltage of the EFLCC changes to 20 kV, and the transmission power $P_{LCC}$ in this scenario remains at 200 MW, while $P_{MMC}$ of the FBMMC changes to -100 MW. After 0.5 seconds (the low-voltage detection time set for the FBMMC part in this paper), which corresponds to $t_{11}$, the EFLCC recalculates the firing angle of the LCC part based on the detected information to ensure that the dc-port voltage of the FBMMC part is maintained at 0 V, enhancing the operational efficiency of the EFLCC during low-power transmission.

It should be noted that after $t_1$, due to the decrease in the dc-voltage-boosting effect of the EFLCC, the dc-port voltage of the rectifier in the LCC-HVDC system will rise accordingly. This phenomenon indicates that the use of the EFLCC can effectively mitigate the dc voltage drop issue in very long-distance HVDC transmission systems, further validating the effectiveness of the approach proposed in Fig. 1. Additionally, both before and after $t_{11}$, the EFLCC operates normally, indicating that for the low-power operating conditions, regardless of whether the LPC is utilized, the EFLCC can transmit the corresponding power as a GFM HVDC tap, although their efficiency may differ at the low power operation.

It is important to note that in the simulation model, the valve-side voltage of the FBMMC part has intentionally been increased to 30 kV, exceeding the conventional 20 kV (40 kV/2). This adjustment is intended to reduce the current in the full-controlled switches of the FBMMC part and ensure a better current-carrying capacity match between the LCC part and the FBMMC part. The successful operation of the simulation validates the correctness of the parameters outlined in Tables I and II. Furthermore, as shown in Fig. 8, regardless of whether the LPC is utilized, the firing angle of the LCC part is calculated in an open-loop manner, leading to no fluctuations. This effectively prevents variations in the firing angle from adversely affecting the safe operation of the EFLCC.

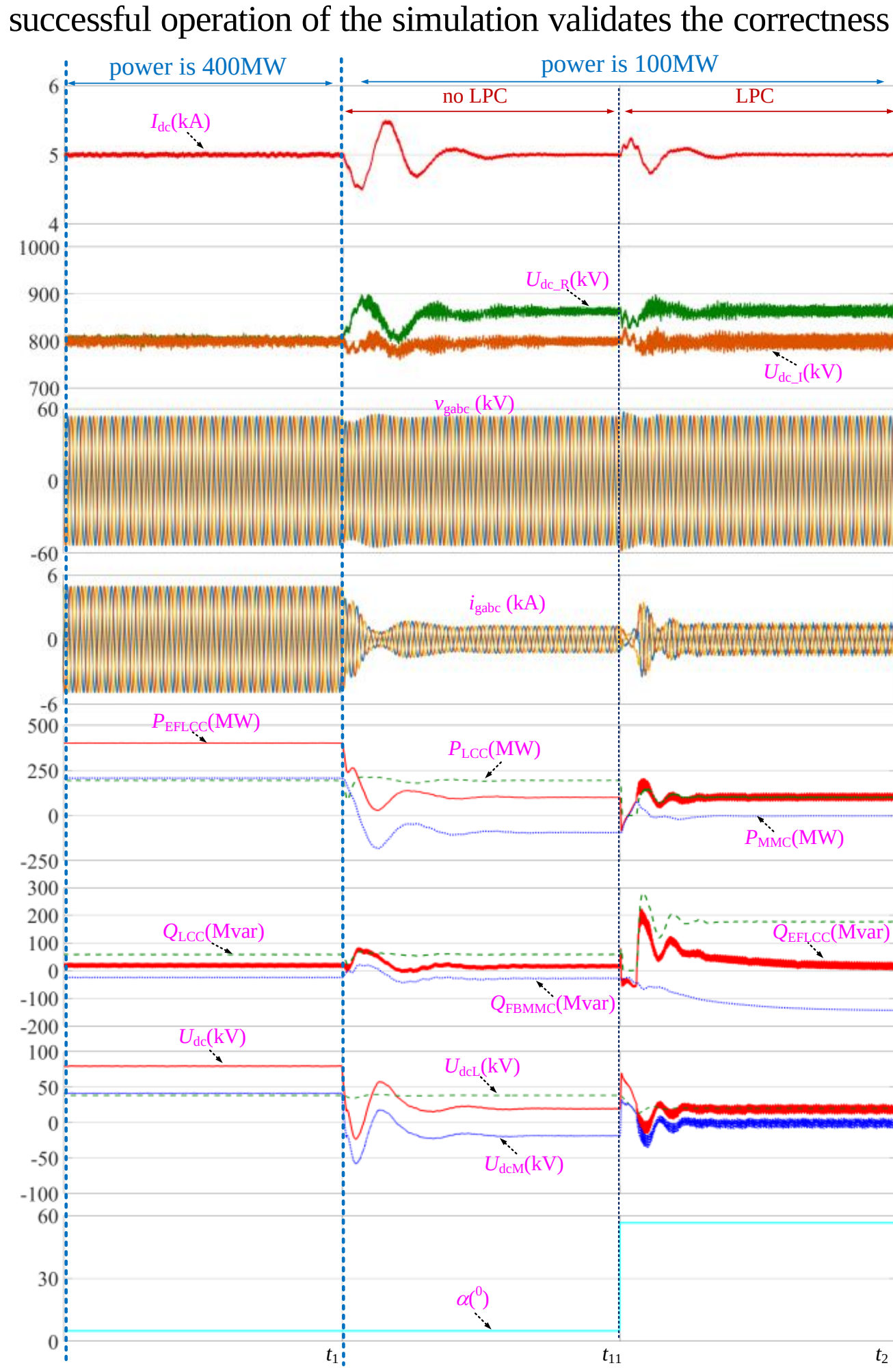

**Fig. 8.** System-level application simulation results of the proposed EFLCC.

## VII. Conclusion

Using the extended controls and parameter designs derived from the Flex-LCC, this paper proposes a new GFM HVDC series tap, the EFLCC. The proposed EFLCC is designed to integrate tens to hundreds of MW of islanded or weak-grid-supported RESs with neighboring LCC-HVDC corridors that can handle thousands of MW. This integration scheme can either increase the proportion of clean energy in the existing HVDC corridor or effectively mitigate the dc voltage drop issue in very long-distance HVDCs, thereby helping to address the two new challenges mentioned in the process of achieving China's dual-carbon goals.

According to the analysis, the capacity of the FBMMC part in the EFLCC is approximately 55% to 60% of the total capacity of the tap, while the active power capacity of the LCC part is 50% of the total capacity of the tap. Meanwhile, the valve-side ac voltages and dc-port voltage of the FBMMC part are decoupled in the EFLCC. Compared to the FBMMC-based HVDC tap, these characteristics facilitate the matching design of the current-carrying capacity between the thyristors and full-controlled switches. For example, in the case study presented in this paper, the valve-side ac voltage of the FBMMC part is increased to 1.5 times its original value, from 20 kV to 30 kV. This allows the full-controlled switches to require a smaller current-carrying capacity (RMS value: reduced from 3.35 kA to 2.75 kA and RMS value: reduced from 5.78 kA to 4.70 kA). Additionally, these characteristics will enable the EFLCC to achieve lower cost and higher rated operational efficiency. For instance, in this paper, the estimated power loss (under rated conditions) is reduced from 4.35 MW to 3.77 MW. Furthermore, from the operational mode perspective at the ac and dc ports, the EFLCC is similar to the FBMMC-based HVDC tap, meaning that the EFLCC is able to provide dc characteristics resembling those of CSCs and ac characteristics resembling those of GFM VSCs, with results validated by experiments and simulations.

Finally, it is important to note that under low power operating conditions (i.e., $P_{\mathrm{EFLCC}} \leq P_{\mathrm{LCC_}\alpha *}$), the EFLCC has two control options: whether to use the proposed LPC. When the LPC is utilized, the EFLCC achieves higher efficiency at low power; however, the control and design method will be more complex compared to the EFLCC without the LPC. Considering that grid developers and operators usually perform rigorous calculations to ensure that the constructed HVDC stations are fully utilized when planning RESs, the low power operating scenarios of the EFLCC may be infrequent. Therefore, the decision of whether to use the LPC in the EFLCC should be based on specific application requirements.

## References

[1] Q. Sun et al., "Flex-LCC: A New Grid-Forming HVDC Rectifier for Collecting Large-Scale Renewable Energy," in *IEEE Transactions on Industrial Electronics*, vol. 71, no. 8, pp. 8808-8818, Aug. 2024.

[2] B. Zhou et al., "Principle and Application of Asynchronous Operation of China Southern Power Grid," in *IEEE Journal of Emerging and Selected Topics in Power Electronics*, vol. 6, no. 3, pp. 1032-1040, Sept. 2018.

[3] A. Tosatto, T. Weckesser and S. Chatzivasileiadis, "Market Integration of HVDC Lines: Internalizing HVDC Losses in Market Clearing," in *IEEE Transactions on Power Systems*, vol. 35, no. 1, pp. 451-461, Jan. 2020.

[4] Y. Xue, X. -P. Zhang and C. Yang, "Series Capacitor Compensated AC Filterless Flexible LCC HVDC With Enhanced Power Transfer Under Unbalanced Faults," in *IEEE Transactions on Power Systems*, vol. 34, no. 4, pp. 3069-3080, July 2019.

[5] K. Strunz, S. Schilling, M. Kuschke and A. Czerwinska, "Fast Contingency Analysis and Control for Overload Mitigation in Integrated High Voltage AC and Multi-Terminal HVDC Grids," in *IEEE Transactions on Power Systems*, vol. 39, no. 4, pp. 5575-5590, July 2024.

[6] Y. Wen, Y. Lu, J. Gou, F. Liu, Q. Tang and R. Wang, "Robust Transmission Expansion Planning of Ultrahigh-Voltage AC–DC Hybrid Grids," in *IEEE Transactions on Industry Applications*, vol. 58, no. 3, pp. 3294-3302, May-June 2022.

[7] A. Nami, J. L. Rodriguez-Amenedo, S. Arnaltes, M. Á. Cardiel-Álvarez and R. A. Baraciarte, "Control of the Parallel Operation of DR-HVDC and VSC-HVDC for Offshore Wind Power Transmission," in *IEEE Transactions on Power Delivery*, vol. 37, no. 3, pp. 1682-1691, June 2022.

[8] Y. Xue, X. -P. Zhang and C. Yang, "AC Filterless Flexible LCC HVDC With Reduced Voltage Rating of Controllable Capacitors," in *IEEE Transactions on Power Systems*, vol. 33, no. 5, pp. 5507-5518, Sept. 2018.

[9] H. Rao et al., "The On-site Verification of Key Technologies for Kunbei-Liuzhou-Longmen Hybrid Multi-terminal Ultra HVDC Project," in *CSEE Journal of Power and Energy Systems*, vol. 8, no. 5, pp. 1281-1289, September 2022.

[10] G. Li and J. Liang, "Modular Multilevel Converters: Recent Applications [History]," in *IEEE Electrification Magazine*, vol. 10, no. 3, pp. 85-92, Sept. 2022.

[11] W. Xiang, P. Meng, and J. Wen, "Development and advances in modular multilevel converter based HVDC projects in China, " in *Front. Eng. Manag.*, vol. 10, no. 1, pp. 183–189, Mar. 2023.

[12] Y. Zhang, C. Zhang, X. Cheng, J. Hao, Y. Qu and W. Wang, "Analysis and Research on Transmission Capacity of ±800 kV Yan-Huai UHVDC," *2020 IEEE 4th Information Technology, Networking, Electronic and Automation Control Conference (ITNEC)*, Chongqing, China, 2020, pp. 1470-1475.

[13] Z. Dai et al., "Study on the Transmission Capability of Dianxibei UHVDC Project and the Method to Enhance It in Early Stage of the Project's Operation," *2018 International Conference on Power System Technology (POWERCON)*, Guangzhou, China, 2018, pp. 2799-2803.

[14] J. Zhang, Q. Zhang, T. Xu, Y. Liu and C. Gao, "Research on Optimal Power Generation Allocation Scheme of DC Transmission Line under the Guidance of Carbon Peak and Carbon Neutralization Targets," *2022 IEEE 5th International Electrical and Energy Conference (CIEEC)*, Nangjing, China, 2022, pp. 4722-4727.

[15] D. S. Carvalho Jr, A. M. Silva, J. H. Almeida, et al., "A second and longer ±800kV DC Bipole completes BeloMonte's integration, " Paper B4–101, CIGRE Session, Paris, 2016

[16] N. Kumar, "Commissioning experience and challenges of world's first 800kV, 6000 MW NER Agra multi-terminal HVDC system," Paper B4–109, CIGRE Session, Paris, 2016.

[17] J. Schipper, S. Sim, Q. Dang, et al., ""Representative modelling of very long HVDC cables," *The 17th International Conference on AC and DC Power Transmission (ACDC 2021)*, Online Conference, 2021, pp. 50-55.

[18] M. Mohaddes, T. An, J. Lu, N. Dhaliwal, M. Szechtman and B. R. Andersen, "Innovative compensation methods for large voltage drop effects in very long HVDC transmission systems," in *CSEE Journal of Power and Energy Systems*, vol. 7, no. 1, pp. 1-8, Jan. 2021.

[19] Z. H. Liu, L. Y. Gao, J. Yu, J. Zhang, and L. C. Lu, et al, "Research work of ±1100kV UHVDC technology," Paper B4–105–2014, CIGRE Session, Paris, 2014.

[20] D. Wu, "Challenges in bringing UHVDC from ±800 kV to higher voltages," Paper B4–116–2018, CIGRE Session, Paris, 2018.

[21] L. Hou, S. Zhang, Y. Wei, B. Zhao and Q. Jiang, "A Dynamic Series Voltage Compensator for the Mitigation of LCC-HVDC Commutation Failure," in *IEEE Transactions on Power Delivery*, vol. 36, no. 6, pp. 3977-3987, Dec. 2021.

[22] C. Guo, Z. Yang, B. Jiang and C. Zhao, "An Evolved Capacitor-Commutated Converter Embedded With Antiparallel Thyristors Based Dual-Directional Full-Bridge Module," in *IEEE Transactions on Power Delivery*, vol. 33, no. 2, pp. 928-937, April 2018.

[23] Y. Xue, X. -P. Zhang and C. Yang, "Elimination of Commutation Failures of LCC HVDC System with Controllable Capacitors," in *IEEE Transactions on Power Systems*, vol. 31, no. 4, pp. 3289-3299, July 2016.

[24] N. Chen, K. Zha, H. Qu, F. Li, Y. Xue and X. -P. Zhang, "Economy Analysis of Flexible LCC-HVDC Systems with Controllable Capacitors," in *CSEE Journal of Power and Energy Systems*, vol. 8, no. 6, pp. 1708-1719, November 2022.

[25] P. Bakas et al., "A review of hybrid topologies combining line-commutated and cascadedfull-bridge converters," in *IEEE Trans. Power Electron.*, vol. 32, no. 10, pp. 7435–7448, Oct. 2017.

[26] H. Xiao et al., "Review of hybrid HVDC systems combining line communicated converter and voltage source converter," *Int. J. Elect. Power Energy Syst.*, vol. 129, pp. 1–9, Jul. 2021.

[27] P. Bakas et al., "Review of hybrid multilevel converter topologies utilizing thyristors for HVDC applications," in *IEEE Trans. Power Electron*., vol. 36, no. 1, pp. 174–190, Jan. 2021.

[28] J. Maneiro, S. Tennakoon and C. Barker, "Scalable shunt connected HVDC tap using the DC transformer concept," *2014 16th European Conference on Power Electronics and Applications*, Lappeenranta, Finland, 2014, pp. 1-10.

[29] E. C. Mathew, R. Sharma and A. Das, "A Modular Resonant DC–DC Converter With High Step-Down Ratio for Tapping Power From HVDC Systems," in *IEEE Transactions on Industrial Electronics*, vol. 68, no. 1, pp. 324-332, Jan. 2021.

[30] E. C. Mathew, R. Sharma and A. Das, "A Fault Tolerant Scheme for Integration of Remote Renewable Energy Sources With HVDC System," in *IEEE Transactions on Power Delivery*, vol. 35, no. 6, pp. 2876-2884, Dec. 2020.

[31] E. C. Mathew and A. Das, "Parallel power tapping from LCC HVDC transmission system with Full Bridge Modular Multilevel Converter," *2020 21st National Power Systems Conference (NPSC)*, Gandhinagar, India, 2020, pp. 1-6.

[32] D. H. R. Suriyaarachchi, C. Karawita and M. Mohaddes, "Applicability of full-bridge and half-bridge MMC for tapping LCC HVDC," *13th IET International Conference on AC and DC Power Transmission (ACDC 2017)*, Manchester, UK, 2017, pp. 1-6.

[33] A. Hartshorne, H. d. T. Mouton and U. K. Madawala, "An investigation into series power tapping options of HVDC transmission lines," *2013 1st International Future Energy Electronics Conference (IFEEC)*, Tainan, Taiwan, 2013, pp. 568-573.

[34] R. E. Torres-Olguin, M. Molinas and T. Undeland, "Offshore Wind Farm Grid Integration by VSC Technology With LCC-Based HVDC Transmission," in *IEEE Transactions on Sustainable Energy*, vol. 3, no. 4, pp. 899-907, Oct. 2012.

[35] N. Parida and A. Das, "Modular Multilevel DC–DC Power Converter Topology With Intermediate Medium Frequency AC Stage for HVDC Tapping," in *IEEE Transactions on Power Electronics*, vol. 36, no. 3, pp. 2783-2792, March 2021.

[36] Q. Sun et al., "Analysis and Experimental Validation of Current-Fed Switched Capacitor-Based Modular DC Transformer," in *IEEE Transactions on Industrial Informatics*, vol. 16, no. 8, pp. 5137-5149, Aug. 2020.

[37] M. Mehrabankhomartash, S. Yin, R. P. Kandula, D. Divan and M. Saeedifard, "Analysis and Design Guidelines of the Isolated Modular Multilevel DC–DC Converter With the Impact of Magnetizing Inductance," in *IEEE Transactions on Industrial Electronics*, vol. 70, no. 12, pp. 11911-11922, Dec. 2023.

[38] M. Aredes, C. Portela and F. C. Machado, "A 25-MW soft-switching HVDC tap for /spl plusmn/500-kV transmission lines," in *IEEE Transactions on Power Delivery*, vol. 19, no. 4, pp. 1835-1842, Oct. 2004.

[39] A. Ismail, M. S. Hamad, A. El Zawawi, et al., “A review of recent HVDC tapping topologies,” in *2016 Eighteenth International Middle East Power Systems Conference (MEPCON)*, Cairo, Egypt, 2016, pp. 765-771.

[40] K.-i. Yamashita, G. Tsukamoto and S. Nishikata, "Steady-State Characteristics of a Line-Commutated Converter-Based High-Voltage Direct Current Transmission System for Series-Connected Wind Power Plants," in *IEEE Transactions on Industry Applications*, vol. 56, no. 4, pp. 3932-3939, July-Aug. 2020.

[41] J. Sau-Bassols, A. Egea-Alvarez, E. Prieto-Araujo and O. Gomis-Bellmunt, "Current Source Converter series tapping of a LCC-HVDC transmission system for integration of offshore wind power plants," *11th IET International Conference on AC and DC Power Transmission*, Birmingham, 2015, pp. 1-7.

[42] Akhil C and S. Maiti, "Series HVDC tapping using modular multilevel current source converter," IECON 2016 - 42nd Annual Conference of the IEEE Industrial Electronics Society, Florence, Italy, 2016, pp. 2569-2574.

[43] D. Hadbi, S. Cornet, R. Denis and O. L. Rhazi, "Assessment of Power Electronics Converters for HVDC series tapping station," *2021 23rd European Conference on Power Electronics and Applications (EPE'21 ECCE Europe)*, Ghent, Belgium, 2021, pp. P.1-P.9.

[44] B. Li, M. Guan, D. Xu, R. Li, G. P. Adam and B. Williams, "A series HVDC power tap using modular multilevel converters," *2016 IEEE Energy Conversion Congress and Exposition (ECCE)*, Milwaukee, WI, USA, 2016, pp. 1-7.

[45] P. Meng, W. Xiang, Y. Chi, Z. Wang, W. Lin, and J. Wen, “Resilient DC voltage control for islanded wind farms integration using cascaded hybrid HVDC system,” *IEEE Trans. Power Syst.*, vol. 37, no. 2, pp. 1054–1066, Feb. 2021.

[46] K. Sun, H. Xiao, J. Pan, and Y. Liu, “A station-hybrid HVDC system structure and control strategies for cross-seam power transmission,” *IEEE Trans. Power Syst.*, vol. 36, no. 1, pp. 379–388, Jan. 2021.

[47] M. Eremia, C. Liu, A. Edris, "CSC–HVDC Transmission," in *Advanced Solutions in Power Systems: HVDC, FACTS, and Artificial Intelligence* , *IEEE*, 2016, pp.35-124.

[48] H. Junxian, Y. Chengyan, L. Yan and L. Chao, "The Adjustment Method and Effect of Reactive Power Control Strategy for LCC-HVDC with Variable Power," 2019 IEEE PES Asia-Pacific Power and Energy Engineering Conference (APPEEC), Macao, China, 2019, pp. 1-5.

[49] Z. Xu, B. Li, M. Guan and D. Xu, "Capacitor voltage balancing control for series MMC tap," IECON 2017 - 43rd Annual Conference of the IEEE Industrial Electronics Society, Beijing, China, 2017, pp. 4488-4493.

[50] R. Li and L. Xu, "A Unidirectional Hybrid HVDC Transmission System Based on Diode Rectifier and Full-Bridge MMC," in IEEE Journal of Emerging and Selected Topics in Power Electronics, vol. 9, no. 6, pp. 6974-6984, Dec. 2021.